\documentclass[a4paper,11pt]{article}
\usepackage{amsmath,eurosym,amssymb,amsfonts,amsfonts,wasysym,latexsym, multirow, multicol,bm,color}
\usepackage{graphicx}
\RequirePackage[numbers,sort&compress]{natbib}

\begin{document}

\renewcommand{\thefootnote}{\fnsymbol{footnote}}
\begin{center}
{\Large \bf Cosmological dynamics of tachyonic teleparallel dark energy}
\vskip 0.5cm
{\bf  G. Otalora}
\vskip 0.3cm
{\it Instituto de F\'{\i}sica Te\'orica, UNESP-Univ Estadual Paulista \\
Caixa Postal 70532-2, 01156-970 S\~ao Paulo, Brazil}

\vskip 0.8cm
\begin{quote}
{\bf Abstract.~}{\footnotesize A detailed dynamical analysis of the tachyonic teleparallel dark energy model, in which a non-canonical scalar field (tachyon field) is non-minimally coupled to gravitation, is performed. It is found that, when the non-minimal coupling is ruled by a dynamically changing coefficient $\alpha\equiv f_{,\phi}/\sqrt{f}$, with $f(\phi)$ an arbitrary function of the scalar field $\phi$, the universe may experience a field-matter-dominated era ``$\phi$MDE'', in which it has some portions of the energy density of $\phi$ in the matter dominated era. This is the most significant difference in relation to the so-called teleparallel dark energy scenario, in which a canonical scalar field (quintessence) is non-minimally coupled to gravitation. 
}
\end{quote}
\end{center}
\vskip 0.8cm

\section{Introduction}
One of the greatest enigmas of modern cosmology is the accelerated expansion of the universe. This result emerges from cosmic observations of Supernovae Ia (SNe Ia) \cite{obs1}, cosmic microwave background (CMB) radiation \cite{obs2}, large scale structure (LSS) \cite{obs3}, baryon acoustic oscillations (BAO) \cite{obs4}, and weak lensing \cite{obs5}.
There are two main approaches to explain such behavior, apart from the simple consideration of a cosmological constant. One is to modify the gravitational sector by generalizing the Einstein-Hilbert action of general relativity (GR), which gives rise to the so-called $F(R)$ theories \cite{2}.
The other approach is based on ``modified matter models'', which consists in introducing an exotic matter source (``dark energy'') with a large negative pressure which is the
dominant fraction of the energy content of the present universe. In this case, the dark energy models can be based on a canonical scalar field (quintessence), or on a non-canonical scalar field (phantom field, tachyon field, k-essence, amongst others) \cite{1,3}. Typically, the scalar field is minimally coupled to gravity, and an explicit coupling of the field to a background fluid
can be implemented or not \cite{4,31}. Also, a non-minimal coupling between the scalar field and gravity is not to be excluded \cite{25,5,6,7,8,9, O. Bertolami2000, Piao2003, Rshidi2013}. Other dark energy models using
covariant versions with non-minimal coupling can also be found in the literature \cite{O. Bertolami2010}.

In analogy to a similar construction in GR, it was proposed in Ref.~\cite{10} a non-minimal coupling between quintessence and gravity in the framework of teleparallel gravity (TG). 
This theory has a rich structure, and has been called ``teleparallel dark energy''; its dynamics was studied later in Refs.~\cite{11,12,13}. 
TG is an alternative description to the geometric description of gravitation (GR). It is a gauge theory for the translation group that is fully equivalent to GR, 
in which the torsionless Levi-Civita connection is replaced by the curvatureless Weitzenb\"{o}ck connection, and the dynamical objects are the four linearly independent tetrads, not the metric tensor \cite{14,15,16,16b}. 
But, despite equivalent to GR, TG is, conceptually speaking, a completely different theory. For example, it attributes gravitation to torsion, which acts as a force, whereas GR attributes gravitation to curvature, which is used to geometrize the gravitational interaction \cite{16b}.
Also, as a gauge theory, TG is closer to the description of the other fundamental interactions, and this can be a conceptual advantage in relation to GR in a possible unification scenario. Furthermore, since its lagrangian depends on the tetrad and on the first derivative of the tetrad, in contrast to GR whose lagrangian depends also on the second derivative of the metric, it turnout to be a simpler theory \cite{16b}.
Now,  when one introduces a scalar field as source of dark energy, in the non-minimal case the additional scalar sector is coupled to the torsion scalar in the TG case, and to the curvature scalar in GR; the resulting coupled equations do not coincide, which implies that the resulting theories are completely different \cite{12,13}.
For the teleparallel gravity generalization, the so-called $F(T)$ theory, see Refs. \cite{1,17,18}.

On the other hand, the tachyon field arising in the context of string theory provides an example of modified form of matter, which has been studied in applications to cosmology both as a source of early inflation and of late-time speed-up of the cosmic expansion rate \cite{19,20,22}. 
The dynamics of the tachyon field is very different from the standard case (quintessence). As the lagrangian of quintessence generalizes the lagrangian of a non-relativistic particle, the lagrangian of the tachyon field generalizes the lagrangian of the relativistic particle \cite{19}.
In this regard the tachyon field generalizes the quintessence field, and a non-minimal version in the context of TG was proposed in Ref.~\cite{23}.

In this paper we will be interested in the dynamics of tachyonic teleparallel dark energy, as this model has been called \cite{23}. Given the nature of the tachyon field, we can expect a richer structure than in the case of teleparallel dark energy. In fact, as we are going to see, an era $\phi$MDE (see Ref. \cite{4}) is possible, but in order to have a viable cosmological evolution it is necessary to
generalize the non-minimal coupling to a dynamically changing coefficient $\alpha\equiv f_{,\phi}/\sqrt{f}$, with $f(\phi)$ the general non-minimal coupling function.

\section{Tachyon field in General Relativity}

The action for the tachyon scalar field minimally coupled with gravity is given by
\begin{equation}
S_{\varphi}=\int d^{4}x\,\sqrt{-g}\,\left[\frac{R}{2\,\kappa^{2}}-V(\varphi)\,\sqrt{1-2\,X}\right],
\label{1}
\end{equation}
where $X=\frac{1}{2}\,\partial_{\mu}\varphi\,\partial^{\mu}\varphi$, $\kappa^{2}=8\,\pi\,G$, and $c=1$ (we adopt natural units and have a metric signature $(+,-,-,-)$). 
$V(\varphi)$ is the potential of the tachyon field, and the potential corresponding to scaling solutions (i.e., the field energy density $\rho_{\varphi}$ is proportional to the fluid energy density $\rho_{m}$) is the inverse
power-law type, $V(\varphi)\propto \varphi^{-2}$. 
Moreover, a remarkable feature of the stress tensor of the tachyon field is that it can be considered as the sum of a pressure-less dust component and a cosmological constant \cite{19}. This means that the stress tensor can be thought of as made up of two components, one behaving like a pressure-less fluid (dark matter), while the other having a negative pressure (dark energy). 
This property is reflected in that when $\dot{\varphi}$ is small compared to unity (compared to $V(\varphi)$ in the case of quintessence), the tachyon field has equation of state $\omega_{\varphi}\rightarrow-1$ and mimic a cosmological constant, just like the quintessence field. But, when $\dot{\varphi}\rightarrow1$ the tachyon field has equation of state $\omega_{\varphi}\approx0$ and behaves like non-relativistic matter with $\rho_{\varphi}\propto a(t)^{-3}$ ($a(t)$ is the scale factor), whereas in the case of quintessence 
for $\dot{\varphi}>>V(\varphi)$, it has equation of state $\omega_{\varphi}\approx1$ (stiff matter) leading to $\rho_{\varphi}\propto a(t)^{-6}$. 
So, the dynamics the tachyon field is very different from the standard field case, irrespective of the steepness of the tachyon potential the equation of state varies between $0$ and $-1$, and the energy density behaves as $\rho_{\varphi}\propto a(t)^{-m}$ with  $0\leq m\leq3$ \cite{1}. 

A study of dynamical systems in Friedmann-Robertson-Walker (FRW) cosmology within phenomenological theories based on the effective tachyon action \eqref{1} can be found in \cite{1,20,22}. In \cite{22}  was proposed perform a transformation of the form
\begin{equation}
\varphi \rightarrow \phi=\int d\varphi \sqrt{V(\varphi)} \Longleftrightarrow \partial \varphi=\frac{\partial \phi}{\sqrt{V(\phi)}}, 
\label{2}
\end{equation}
which allows to introduce normalized phase-space variables and in terms of these variables one can obtain a closed autonomous system of ordinary differential equations (ODE) out of the cosmological field equations written in terms of the transformed tachyon field $\phi$, for a broad class of self-interaction potentials $V(\phi)$ (in \cite{31} also was carried out a transformation  of this type to study coupled dark energy in GR).
Also, as we will show quite soon, the above field re-definition allows us to study a non-minimal coupling between tachyon field and teleparallel gravity in terms a closed autonomous system of ODE. We are going to concentrate on the inverse square potential $V(\varphi)\propto \varphi^{-2}$,  that for the transformed field $\phi$ becomes $V(\phi)=V_{0}\,e^{-\lambda\,\kappa\,\phi}$, and $\lambda$ is a constant.

\section{Tachyonic teleparallel dark energy}

In what follows we consider a non-minimal coupling between tachyon field and teleparallel gravity as was already considered in Ref. \cite{23}. In order to have a closed autonomous system of ODE and study the dynamics of the model is required the transformation $\varphi \rightarrow \phi$ in accordance to \eqref{2}. 
Under the transformation \eqref{2}, the relevant action reads
\begin{equation}
 S=\int d^{4}x\,h\,\left[\frac{T}{2\,\kappa^2}-V(\phi)\,\sqrt{1-\frac{2\,X}{V(\phi)}}+\xi\,f(\phi)\,T\right]+S_{m},
 \label{3}
\end{equation} where $h\equiv\det(h^{a}_{~\mu})=\sqrt{-g}$ ( $h^{a}_{~\mu}$ are the orthonormal components of the tetrad), $T/2\,\kappa^2$ is the lagrangian of teleparallelism ($T$ is the torsion scalar), $S_{m}$ is the matter action, $\xi$ is a dimensionless constant measuring the non-minimal coupling, and $f(\phi)>0$ is the non-minimal coupling function with units of $mass^{2}$ that only depends of the transformed tachyon field $\phi$ (see Refs. \cite{13,16b}). 
Varying the action \eqref{3} with respect to tetrad fields yields field equation
\begin{multline}
2\,\left(\frac{1}{\kappa^{2}}+2\,\xi\,f(\phi)\right)\left[h^{-1}\,h^{a}_{~\alpha}\,\partial_{\sigma}\left(h\,h_{a}^{~\tau}\,S_{\tau}^{~\rho\sigma}\right)+T^{\tau}_{~\nu\alpha}\,S_{\tau}^{~\rho\nu}+\frac{T}{4}\,\delta^{~\rho}_{\alpha}\right]\\
+4\,\xi\,S_{\alpha}^{~\rho\sigma}\,f_{,\phi}\,\partial_{\sigma}\phi-\mu^{-1}\,V(\phi)\,\delta^{~\rho}_{\alpha}-\mu\,\partial_{\alpha}\phi\,\partial^{\rho}\phi=\Theta_{\alpha}^{~\rho}.
 \label{4}
\end{multline}where $\Theta_{\alpha}^{~\rho}$ stands for the symmetric energy-momentum tensor, $T^{\tau}_{~\nu\alpha}$ is the torsion tensor and $S_{\tau}^{~\rho\sigma}$ is the superpotential (see Ref. \cite{16b}). Also, we define $f_{,\phi}\equiv \frac{d f(\phi)}{d\phi}$ and
\begin{equation}
 \mu\equiv\frac{1}{\sqrt{1-\frac{2\,X}{V}}}.
 \label{5}
\end{equation} Imposing the flat FRW geometry (see Ref. \cite{10}),
\begin{equation}
 h^{a}_{~\mu}(t)= \mbox{diag}(1,a(t),a(t),a(t)),
 \label{6}
\end{equation} we obtain the Friedmann equations with
\begin{equation}
 \rho_{\phi}=\mu\,V\left(\phi\right)-6\,\xi\,{H}^{2}\,f(\phi),
 \label{7}
\end{equation} the scalar energy density and
\begin{equation}
 p_{\phi}=-\mu^{-1}\,V(\phi)+4\,\xi\,H\,f_{,\phi}\,\dot{\phi}+2\,\xi\,\left(3\,H^2+2\,\dot{H}\right)\,f(\phi),
 \label{8}
\end{equation}the pressure density of field. Here we also use the useful relation $T=-6\,H^{2}$, which arises for flat FRW geometry.

Also, in the flat FRW background, the variation of the action \eqref{3} with respect to scalar field yields the motion equation

\begin{equation}
 \ddot{\phi}+3\,\mu^{-2}\,H\,\dot{\phi}+\left(1- \frac{3\,X}{V}\right)V_{,\phi}+6\,\xi\,\mu^{-3}\,f_{,\phi}\,H^2=0.
 \label{9}
\end{equation} Rewriting the equation of motion \eqref{9} in terms of scalar energy density and the pressure density of field we obtain
\begin{equation}
 \dot{\rho}_{\phi}+3\,H\,\rho_{\phi}\left(1+\omega_{\phi}\right)=0,
\label{10}
\end{equation}whereas that for matter
\begin{equation}
  \dot{\rho}_{m}+3\,H\,\rho_{m}\left(1+\omega_{m}\right)=0,
 \label{11}
\end{equation}where $\omega_{\phi}\equiv p_{\phi}/\rho_{\phi}$ and $\omega_{m}\equiv p_{m}/\rho_{m}=const$ are the equation-of-state parameter of dark energy and dark matter, respectively. We also define the barotropic index $\gamma\equiv1+\omega_{m}$, such that $0<\gamma<2$. 
On the other hand, we note that there is no coupling between dark energy and dark matter.

\section{Phase-space analysis}
In order to study the dynamics of the model it is convenient to introduce the following dimensionless
variables
\begin{equation}
 x\equiv\frac{\dot{\phi}}{\sqrt{V}}, \:\:\:\:\:\: y\equiv\frac{\kappa\,\sqrt{V}}{\sqrt{3}\,H}, \:\:\:\:\:\:  u\equiv\kappa\,\sqrt{f}, \:\:\:\:\ \alpha\equiv\frac{f_{,\phi}}{\sqrt{f}}, \:\:\:\: \lambda\equiv-\frac{V_{,\phi}}{\kappa V}.
 \label{12}
\end{equation} Using these variables we define
\begin{equation}
 s\equiv-\frac{\dot{H}}{H^{2}}=\frac{4\,\sqrt{3}\,\alpha\,\xi\,u\,x\,y+3\,\mu\left( {x}^{2}-\gamma\right) \,{y}^{2}}{2\,\left( 2\,\xi\,{u}^{2}+1\right)}+\frac{3\,\gamma}{2}.
 \label{13}
\end{equation}Also, using \eqref{12} the evolution equations \eqref{10} and \eqref{11} can be rewritten as a dynamical system of ODE, namely
\begin{equation}
 x'=\frac{\sqrt{3}}{2}\,\left(\lambda\,{x}^{2}\,y+\lambda\,\left( 2-3\,{x}^{2}\right) \,y-4\,\alpha\,\xi\,u\,\mu^{-3}\,y^{-1}-2\,\sqrt{3}\,x\,\mu^{-2}\right),
 \label{14}
 \end{equation}
\begin{equation}
 y'=\left(-\frac{\sqrt{3}\,\lambda}{2}\,x\,y+s\right)\,y,
 \label{15}
\end{equation}

\begin{equation}
 u'=\frac{\sqrt{3}\,\alpha\,x\,y}{2},
 \label{16}
\end{equation}

\begin{equation}
 \lambda'=-\sqrt{3}\,\lambda^{2}\,x\,y\,\left(\Gamma-1\right),
 \label{17}
\end{equation}

\begin{equation}
 \alpha'=\sqrt{3}\,\frac{\,x\,y\,\alpha^{2}}{u}\left(\Pi-\frac{1}{2}\right),
 \label{18}
\end{equation}with $\mu=1/\sqrt{1-x^{2}}$ and prime denotes  derivative with respect to the so-called e-folding  time $N\equiv\ln{a}$. Also,  we define

\begin{equation}
\Pi\equiv\frac{f\,f_{,\phi\phi}}{f_{,\phi}^{2}}, \:\:\:\:\: \:\:\:\:\:\: \Gamma\equiv\frac{V\,V_{,\phi\phi}}{V_{,\phi}^2}.
\label{19}
\end{equation}
The fractional energy densities $\Omega\equiv(\kappa^{2}\,\rho)/(3\,H^{2})$ for the scalar field and background matter are given by
\begin{equation}
 \Omega_{\phi}=\mu\,{y}^{2}-2\,\xi\,{u}^{2}, \:\:\:\:\:\:\:\: \Omega_{m}=1-\Omega_{\phi}.
 \label{20}
\end{equation}
The state equation of the field $\omega_{\phi}=p_{\phi}/\rho_{\phi}$ reads
\begin{equation}
 \omega_{\phi}=\frac{-\,\mu^{-1}\,{y}^{2}+2\,\xi\,u\,\left( \frac{2\,\sqrt{3}}{3}\,\alpha\,x\,y+u\,\left(1-\frac{2}{3}\,s\right)\right)}{\mu\,{y}^{2}-2\,\xi\,{u}^{2}}.
 \label{21}
\end{equation}
On the other hand, the effective equation of state $\omega_{eff}=\left(p_{\phi}+p_{m}\right)/\left(\rho_{\phi}+\rho_{m}\right)$ is given by
\begin{equation}
\omega_{eff}=\left({x}^{2}-\gamma\right)\mu \,{y}^{2}+ \frac{4\,\sqrt{3}}{3}\,\alpha\,\xi\,u\,x\,y+2\,\left(\gamma-\frac{2}{3}\,s\right)\,\xi\,{u}^{2}+\,\gamma-1,
\label{22}
\end{equation}and the accelerated expansion of the universe occurs for $\omega_{eff}<-1/3$. 

Once the parameters $\Gamma$ and $\Pi$ are known, the dynamical system \eqref{14}-\eqref{18} becomes an autonomous system and the dynamics can be analyzed in the usual way.
Since we consider constant $\lambda$, this is equivalent to consider $\Gamma=1$. On the other hand, for $f(\phi)\propto\phi^{2}$ or equivalently $\Pi=1/2$ then $\alpha\equiv f_{,\phi}/\sqrt{f}=const\neq0$. 
Moreover, following Ref. \cite{13},  for a general coupling function $u\equiv\kappa\,\sqrt{f(\phi)}$, with inverse function $\phi=f^{-1}(u^{2}/\kappa^{2})$,  $\alpha(\phi)$ and $\Pi(\phi)$  can be expressed in terms of $u$ (this approach is similar to that followed in the case of quintessence in GR with potential beyond exponential potential \cite{30}). 
Therefore, two situations may arise; one where $\alpha$ is a constant and  another where $\alpha$ depends on $u$. In both cases, we have a three-dimensional autonomous system \eqref{14}-\eqref{16}, and the fixed points or critical points $(x_{c},y_{c},u_{c})$ can be find by imposing the conditions $x'_{c}=y'_{c}=u'_{c}=0$. 
From the definition \eqref{12}, $x_{c}$, $y_{c}$, $u_{c}$  should be real, with $x_{c}^{2}\leq1$, $y_{c}\geq0$, and $u_{c}\geq0$. 

To study the stability of the critical point, we substitute linear perturbations, $x\rightarrow x_{c}+\delta{x}$, $y\rightarrow y_{c}+\delta{y}$, and $u\rightarrow u_{c}+\delta{u}$  about the critical point $(x_{c},y_{c},u_{c})$ into the autonomous system \eqref{14}-\eqref{16} and linearize them.
The eigenvalues of the perturbations matrix $\mathcal{M}$, namely, $\tau_{1}$, $\tau_{2}$ and $\tau_{3}$,  determine the conditions of stability of the critical points. One generally uses the following
classification (see Refs. \cite{1,3}):
(i) Stable node: $\tau_{1}<0$, $\tau_{2}<0$ and $\tau_{3}<0$. (ii) Unstable node: $\tau_{1}>0$, $\tau_{2}>0$ and $\tau_{3}>0$. (iii) Saddle point: one or two of the three eigenvalues are positive and the other negative. (iv) Stable spiral: The determinant of the matrix
$\mathcal{M}$ is negative and the real parts of $\tau_{1}$, $\tau_{2}$ and $\tau_{3}$ are negative.  A  critical point is an attractor in the cases (i) and (iv), but it is not so in the cases (ii) and (iii). The universe will eventually enter these
attractor solutions regardless of the initial conditions.
In what follows we are going to study the three-dimensional autonomous dynamical system \eqref{14}-\eqref{16}, first for  $\alpha=const\neq0$ and  then for  dynamically changing $\alpha(u)$, such that $\alpha(u)\rightarrow\alpha(u_{c})=0$  when the  system falls into the critical point $(x_{c},y_{c},u_{c})$. 

\section{Constant $\alpha$}

\subsection{Critical points}

In this section we consider a non-minimal coupling function $f(\phi)\propto\phi^{2}$ such that $\alpha=const\neq0$. The critical points of the autonomous system \eqref{14}-\eqref{16} are presented in Table 1. 
In Table 2 we summarize the stability properties (to be studied below), and conditions for acceleration and existence for each point.
In Table 1 the variables $v_{\pm}$ are defined by
\begin{equation}
 v_{\pm}=\left(\alpha\,\xi\pm\sqrt{\xi\,\left( {\alpha}^{2}\,\xi-2\,{\lambda}^{2}\right) }\right).
 \label{23}
\end{equation}
The critical point I.a is a fluid dominant solution ($\Omega_{m}=1$) that exists for all values of $\lambda$, $\xi$ and $\alpha$.
The critical points I.b and I.c  are both scaling solutions with $u_{c}\geq0$, and the requirement of the condition $0<\Omega_{\phi}<1$ implies $0<\hat{\xi}<1$. The accelerated expansion occurs for these three points if $\omega_{eff}=\gamma-1<-1/3$, that is, for $\gamma<2/3$.
Points I.d and I.e both correspond to dark-energy-dominated de Sitter solutions  with $\Omega_{\phi}=1$ and $\omega_{\phi}=\omega_{eff}=-1$. From \eqref{23}, the fixed point I.d exists for:
\begin{equation}
 \xi\geq 2\,{\lambda}^{2}/\alpha^{2}>0\:\:\:\: \text{and}\:\:\:\:\lambda/\alpha>0\:\:\:\: \text{or} \:\:\:\:\:\xi<0,\:\:\:\:\alpha<0\:\:\:\: \text{and} \:\:\:\:\lambda>0.
 \label{24}
\end{equation}

By the other hand, the point I.e exists for 
\begin{equation}
 \xi\geq 2\,{\lambda}^{2}/\alpha^{2}>0\:\:\:\: \text{and}\:\:\:\: \lambda/\alpha>0\:\:\:\: \text{or}\:\:\:\: \xi<0,\:\:\:\:\alpha>0\:\:\:\: \text{and}\:\:\:\:\lambda<0.
 \label{25}
\end{equation}

\begin{table}[t]
\caption{Critical points for the autonomous system \eqref{14}-\eqref{16} for constant $\alpha\neq0$. We define $\hat{\xi}\equiv 1+2\,\xi\,u_{c}^{2}$ and $u_{c}\geq0$.}
 \centering
\begin{center}
\begin{tabular}{c c c c c c c}\hline
Name & $x_{c}$ & $y_{c}$ & $u_{c}$  & $\Omega_{\phi}$ & $\omega_{\phi}$ &$\omega_{eff}$\\\hline
I.a & $0$ & $0$ & $0$ &  $0$ & $-1$ & $\gamma-1$\\\hline
I.b & $1$ & $0$ & $u_{c}$ &  $1-\hat{\xi}$ & $\gamma-1$ & $\gamma-1$\\\hline
I.c & $-1$ & $0$ & $u_{c}$ &  $1-\hat{\xi}$ & $\gamma-1$ & $\gamma-1$\\\hline
I.d & $0$ & $\sqrt{\frac{\alpha\,v_{-}}{\lambda^{2}}}$ & $\frac{v_{-}}{2\,\lambda\,\xi}$ & $1$ & $-1$ & $-1$\\\hline
I.e & $0$ & $\sqrt{\frac{\alpha\,v_{+}}{\lambda^{2}}}$ & $\frac{v_{+}}{2\,\lambda\,\xi}$ & $1$ & $-1$ & $-1$\\\hline
\end{tabular}
\end{center}
\end{table}

\begin{table}[t]
\caption{Stability properties, and conditions for acceleration and existence of the fixed points in Table 1.}
 \centering
\begin{center}
\begin{tabular}{c c c c c c c}\hline
Name & Stability & Acceleration & Existence \\\hline
I.a &  Unstable &    $\gamma<2/3$       &   All values  \\\hline
I.b &  Saddle  &    $\gamma<2/3$        &   $0<\hat{\xi}<1$ \\\hline
I.c &  Saddle  &    $\gamma<2/3$     & $0<\hat{\xi}<1$ \\\hline
I.d &  Stable node or stable spiral, or saddle &  All values  & Eq. \eqref{24} \\\hline
I.e &  Stable node or stable spiral, or saddle   & All values   & Eq. \eqref{25} \\\hline
\end{tabular}
\end{center}
\end{table}

\subsection{Stability}

Substituting the  linear perturbations, $x\rightarrow x_{c}+\delta{x}$, $y\rightarrow y_{c}+\delta{y}$, and $u\rightarrow u_{c}+\delta{u}$  into the autonomous system \eqref{14}-\eqref{16} and linearize them, the components of the matrix of perturbations $\mathcal{M}$ are given by

\begin{equation}
\mathcal{M}_{11}=\sqrt{3}\,\left( -2\,\lambda\,x_{c}\,y_{c}-\sqrt{3}\left(1-3\,x^{2}_{c}\right)+6\,\alpha\,\xi\,u_{c}\,x_{c}\,\mu^{-1}_{c}\, y^{-1}_{c}\right),
\label{26}
\end{equation}

\begin{equation}
 \mathcal{M}_{12}=\sqrt{3}\,\mu^{-2}_{c}\,\left( \lambda+2\,\alpha\,\xi\,u_{c}\,\mu^{-1}_{c}\,y_{c}^{-2}\right),
 \label{27}
\end{equation}

\begin{equation}
 \mathcal{M}_{13}=-2\,\sqrt{3}\,\alpha\,\xi\,\mu^{-3}_{c}\,y^{-1}_{c},
 \label{28}
\end{equation}

\begin{equation}
 \mathcal{M}_{21}=\frac{\,{y}^{2}_{c}\,\left(-3\,\mu^{3}_{c}\,x_{c}\,\,y_{c}\,\left( {x}^{2}_{c}+\gamma-2\right)+4\,\sqrt{3}\,\alpha\,\xi\,u_{c}\right)}{2\,\left( 2\,\xi\,{u}^{2}_{c}+1\right)}-\frac{\sqrt{3}\,\lambda\,y_{c}^{2}}{2},
 \label{29}
 \end{equation}

 \begin{equation}
\mathcal{M}_{22}=\frac{y_{c}\left(9\,\mu_{c}\,{y}_{c}\,\left( {x}^{2}_{c}-\gamma\right)+8\,\sqrt{3}\,\alpha\,\xi\,x_{c}\,u_{c}\right)}{2\,\left( 2\,\xi\,{u}^{2}_{c}+1\right)}-\sqrt{3}\,\lambda\,x_{c}\,y_{c}+\frac{3\,\gamma}{2},
\label{30}
 \end{equation}
 
 \begin{equation}
  \mathcal{M}_{23}=\frac{2\,\sqrt{3}\,\xi\,{y}^{2}_{c}\,\left(-\sqrt{3}\,\mu_{c}\,u_{c}\,y_{c}\,\left( {x}^{2}_{c}-\gamma\right)+2\,\alpha\,x_{c}\right) }{{\left( 2\,\xi\,{u}^{2}_{c}+1\right) }^{2}}-\frac{2\,\sqrt{3}\,\alpha\,\xi\,x_{c}\,y_{c}^{2}}{2\,\xi\,u_{c}^{2}+1},
  \label{31}
 \end{equation}
 
 \begin{equation}
 \mathcal{M}_{31}=\frac{\sqrt{3}\,\alpha\,y_{c}}{2}, \:\:\:\:\:\: \mathcal{M}_{32}=\frac{\sqrt{3}\,\alpha\,x_{c}}{2}, \:\:\:\:\:\: \mathcal{M}_{33}=0.
 \label{32}
 \end{equation}In the above we define $\mu_{c}=1/\sqrt{1-x_{c}^{2}}$.

Point I.a:
The component $\mathcal{M}_{13}$ is divergent, which means that this point is unstable. 

Points I.b and I.c:
For both  critical points the eigenvalues of $\mathcal{M}$ are given by
\begin{equation}
 \tau_{1}=\frac{3\,\gamma}{2}, \:\:\:\:\: \tau_{2}=-3, \:\:\:\:\:\:\: \tau_{3}=0. 
 \label{33}
\end{equation}Therefore these points are unstable.

Point I.d:
For this point the eigenvalues are given by
\begin{equation}
 \tau_{1,2}=\frac{3\,\left(-1\pm\sqrt{1-\frac{4}{3}\,\alpha\,\sqrt{\xi\,\left( {\alpha}^{2}\,\xi-2\,{\lambda}^{2}\right)}}\right)}{2}, \:\:\:\:\:\:\: \tau_{3}=-3\,\gamma.
 \label{34}
\end{equation}
It is a saddle point if $\xi<0$ and $\alpha<0$ or $\xi>2\,\lambda^{2}/\alpha^{2}$ and $\alpha<0$. 
On the other hand, for
\begin{equation}
 \frac{2\,\lambda^{2}}{\alpha^{2}}<\xi\leq\frac{\lambda^{2}}{\alpha^{2}}\,\left(1+\sqrt{1+\frac{9}{16\,\lambda^{4}}}\right),
 \label{35}
\end{equation}and $\alpha>0$ it is a stable node. Also, when $\xi>\frac{\lambda^{2}}{\alpha^{2}}\,\left(1+\sqrt{1+\frac{9}{16\,\lambda^{4}}}\right)$ then $\tau_{1}$ and $\tau_{2}$ are complex with real part negative and 
$\det(\mathcal{M})=-9\,\alpha\,\gamma\,\sqrt{\xi\,\left( {\alpha}^{2}\,\xi-2\,{\lambda}^{2}\right) }<0$  for $\alpha>0$. Therefore, in this case it is a stable spiral.

Point I.e:
Finally, for the point I.e, the eigenvalues are 
\begin{equation}
  \tau_{1,2}=\frac{3\,\left(-1\pm\sqrt{1+\frac{4}{3}\,\alpha\,\sqrt{\xi\,\left( {\alpha}^{2}\,\xi-2\,{\lambda}^{2}\right) }}\right)}{2}, \:\:\:\:\:\:\: \tau_{3}=-3\,\gamma.
  \label{36}
\end{equation}

This fixed point is a saddle point for $\xi<0$ and $\alpha>0$ or $\xi>2\,{\lambda}^{2}/\alpha^{2}$ and $\alpha>0$. On the other hand, for $\xi$ as in \eqref{35} and $\alpha<0$, it is a stable node.
When $\xi>\frac{\lambda^{2}}{\alpha^{2}}\,\left(1+\sqrt{1+\frac{9}{16\,\lambda^{4}}}\right)$ then $\tau_{1}$ and $\tau_{2}$ are complex with real part negative and  $\det(\mathcal{M})=9\,\alpha\,\gamma\,\sqrt{\xi\,\left( {\alpha}^{2}\,\xi-2\,{\lambda}^{2}\right) }<0$  for $\alpha<0$. In this case, point I.e is a stable spiral.

The fixed points I.a, I.d and I.e are the same points that were found for teleparallel dark energy in Ref. \cite{11,12,13}. The scaling solutions I.b and I.c are new solutions that are not present in teleparallel dark energy.
Such as in teleparallel dark energy, in tachyonic teleparallel dark energy the universe is attracted for the dark-energy-dominated de Sitter solution I.d or I.e. However, unlike the former scenario, in tachyonic teleparallel dark energy the universe  may present a phase $\phi$MDE, that is, the scaling solution I.b or I.c,
in which it has some portions of the energy density of $\phi$ in the matter dominated era. This type of phase $\phi$MDE is also common in coupled dark energy in GR (see Refs. \cite{1, 4, 31}).
But since the scaling solutions I.b and I.c both require $-1/2\,u_{c}^2<\xi<0$ when $u_{c}>0$, then the fixed points I.d and I.e are not achieved because in this case these are saddle points. To solve this problem is necessary to consider a dynamically changing $\alpha$. 

\section{Dynamically changing $\alpha$}

Following Ref. \cite{13}, now we let us consider a general function of non-minimal coupling  $f(\phi)$ such that $\alpha$ can be expressed in terms of $u$ and $\alpha(u)\rightarrow\alpha(u_{c})=0$  when $(x,y,u)\rightarrow (x_{c},y_{c},u_{c})$   (we note that $(x_{c},y_{c},u_{c})$ is a fixed point of the system).
The field $\phi$ rolls down toward $\pm \infty$ ($x>0$ or $x<0$) with $f(\phi)\rightarrow u_{c}^{2}/\kappa^{2}$  when $(x,y,u)\rightarrow (x_{c},y_{c},u_{c})$ (for simplicity and since we seek new solutions then we set $x_{c}\neq0$ and $y_{c}\neq0$). The fixed points are presented in Table 3. Also, we summarize the properties of the fixed points in Table 4. In Table 3 the parameter $y_{c}$ is defined by
\begin{equation}
 y_{c}=\sqrt{\frac{\hat{\xi}\,\left(-{\lambda}^{2}\,\hat{\xi}+\sqrt{{\lambda}^{4}\,{\hat{\xi}}^{2}+36}\right)}{6}}.
 \label{37}
\end{equation}
\subsection{Critical points}
Points II.a and II.b are scaling solutions in which the energy density of the scalar field decreases proportionally to that of the perfect fluid ($\omega_{\phi}=\omega_{m}$). The existence of these solutions 
requires the condition $0<\gamma<1$ or equivalently $-1<\omega_{m}<0$ as can be seen in the expression of $x_{c}$, $y_{c}$ and $\Omega_{\phi}$. Also, for point II.a is required $\lambda<0$ and for point II.b is required $\lambda>0$.   For both points, if $0<\gamma<1$, the condition $0<\Omega_{\phi}<1$ is ensured if
\begin{equation}
\frac{3\,\gamma}{{\lambda}^{2}\,\sqrt{1-\gamma}}<\hat{\xi}<\frac{3\,\gamma}{{\lambda}^{2}\,\sqrt{1-\gamma}}+1.
\label{38}
\end{equation}The condition for accelerated expansion corresponds to $\gamma<2/3$.

Point II.c  is a scalar-field dominant solution ($\Omega_{\phi}=1$) that gives an accelerated expansion at late times for $\lambda^{2}\,y_{c}^{2}<2$, or equivalently, this condition translates into 
\begin{equation}
0<\hat{\xi}<\frac{2\,\sqrt{3}}{{\lambda}^{2}}.
\label{39}
\end{equation}This point exists for $\hat{\xi}>0$ and all values ​​of $\lambda$.

\begin{table}[t]
 \centering
 \caption{Critical points of the autonomous system \eqref{14}-\eqref{16} for dynamically changing $\alpha(u)$ such that $\alpha(u)\rightarrow\alpha(u_{c})=0$ and $u_{c}\geq0$. We define $\hat{\xi}\equiv 2\,\xi\,u_{c}^{2}+1$. }
\begin{center}
\begin{tabular}{c c c c c c c}\hline
Name & $x_{c}$ & $y_{c}$ & $u_{c}$  & $\Omega_{\phi}$ & $\omega_{\phi}$ &$\omega_{eff}$\\\hline
II.a & $-\sqrt{\gamma}$ & $-\frac{\sqrt{3}\,\sqrt{\gamma}}{\lambda}$ &  $u_{c}$  &  $\frac{3\,\gamma}{{\lambda}^{2}\,\sqrt{1-\gamma}}+1-\hat{\xi}$ & $\gamma-1$ & $\gamma-1$\\\hline
II.b & $\sqrt{\gamma}$ &  $\frac{\sqrt{3}\,\sqrt{\gamma}}{\lambda}$ &  $u_{c}$   &   $\frac{3\,\gamma}{{\lambda}^{2}\,\sqrt{1-\gamma}}+1-\hat{\xi}$ & $\gamma-1$ & $\gamma-1$\\\hline
II.c & $\frac{\lambda\,y_{c}}{\sqrt{3}}$  & $y_{c}$  &   $u_{c}$    &   $1$ & $\frac{{\lambda}^{2}\,y_{c}^{2}}{3}-1$ & $\frac{{\lambda}^{2}\,y_{c}^{2}}{3}-1$ \\\hline
\end{tabular}
\end{center}
\end{table}

\begin{table}[t]
\caption{Stability properties, and conditions for acceleration and existence of the fixed points in Table 3.}
 \centering
\begin{center}
\begin{tabular}{c c c c c c c}\hline
Name & Stability & Acceleration & Existence \\\hline
II.a &  Stable node or stable spiral  &    $\gamma<2/3$        &  Eq. \eqref{38} and $\lambda<0$ \\\hline
II.b &  Stable node or stable spiral   &    $\gamma<2/3$     & Eq. \eqref{38}  and $\lambda>0$ \\\hline
II.c & Stable node  &   $\hat{\xi}<\frac{2\,\sqrt{3}}{{\lambda}^{2}}$   & $\hat{\xi}>0$ \\\hline
\end{tabular}
\end{center}
\end{table}

\subsection{Stability}

For dynamically changing $\alpha(u)$ such that $\alpha(u)\rightarrow\alpha(u_{c})=0$, the components of the matrix of perturbation $\mathcal{M}$ are written as

\begin{equation}
\mathcal{M}_{11}=\sqrt{3}\,\left(-2\,\lambda\,x_{c}\,y_{c}+\sqrt{3}\left(3\,x_{c}^{2}-1\right)\right),
\label{47}
\end{equation}

\begin{equation}
 \mathcal{M}_{12}=\sqrt{3}\,\lambda\,\mu_{c}^{-2},
 \label{48}
\end{equation}

\begin{equation}
 \mathcal{M}_{13}=-2\,\sqrt{3}\,\xi\,\eta_{c}\,u_{c}\,\mu_{c}^{-3}\,y_{c}^{-1},
 \label{49}
\end{equation}

\begin{equation}
 \mathcal{M}_{21}=-\frac{3\,\mu_{c}^{3}\,x_{c}\,y_{c}^{3}\,\left( x_{c}^{2}+\gamma-2\right)}{2\,\left( 2\,\xi\,u_{c}^{2}+1\right) }-\frac{\sqrt{3}\,\lambda\,y_{c}^{2}}{2},
 \label{50}
 \end{equation}

 \begin{equation}
\mathcal{M}_{22}=\frac{9\,\mu_{c}\,\left( x_{c}^{2}-\gamma\right) \,y_{c}^{2}}{2\,\left( 2\,\xi\,u_{c}^{2}+1\right) }-\sqrt{3}\,\lambda\,x_{c}\,y_{c}+\frac{3\,\gamma}{2},
\label{51}
 \end{equation}
 
 \begin{equation}
  \mathcal{M}_{23}=-\frac{6\,\xi\,\mu_{c}\,u_{c}\,y_{c}^{3}\,\left( x_{c}^{2}-\gamma\right)}{{\left( 2\,\xi\,u_{c}^{2}+1\right) }^{2}}+\frac{2\,\sqrt{3}\,\xi\,\eta_{c}\,x_{c}\,y_{c}^{2}\,u_{c}}{2\,\xi\,u_{c}^{2}+1},
  \label{52}
 \end{equation}
 
 \begin{equation}
 \mathcal{M}_{31}=0, \:\:\:\:\:\: \mathcal{M}_{32}=0, \:\:\:\:\:\: \mathcal{M}_{33}=\frac{\sqrt{3}\,\eta_{c}\,x_{c}\,y_{c}}{2}.
 \label{53}
 \end{equation}
 
 Here $\eta_{c}$ is defined by $\eta_{c}\equiv\frac{d\alpha(u)}{du}|_{u=u_{c}}$.
 
 Points II.a and II.b: The eigenvalues are
 \begin{equation}
 \tau_{1,2}=\frac{3\,\left( \pm\sqrt{{\left( 2-\gamma\right) }^{2}+\frac{16\,\gamma\,\left(1-\gamma\right)}{\hat{\xi}}\left(\frac{3\,\gamma}{{\lambda}^{2}\,\sqrt{1-\gamma}}-\hat{\xi}\right)}-\left(2-\gamma\right)\right) }{4},\:\:\:\:\:\:\: \tau_{3}=\frac{3\,\eta_{c}\,\gamma}{2\,\lambda}.
 \end{equation}Both points are stable node or stable spiral provided that $\Omega_{\phi}<1$ and $\eta_{c}>0$ (point II.a) or $\eta_{c}<0$ (point II.b).
In any case, both scaling solutions are not realistic solutions in applying to dark energy because of the condition $\gamma<1$ or equivalently $\omega_{m}<0$. This problem can be solved by considering a explicit coupling to dark matter. In this case, as was shown in Ref. \cite{13} for interacting teleparallel dark energy, scaling attractors with accelerated expansion can be solutions of the system.

Point II.c: The eigenvalues are
\begin{equation}
 \tau_{1}=-3+\frac{{\lambda}^{2}\,{y}_{c}^{2}}{2},\:\:\:\:\:\:\: \tau_{2}=-3\,\gamma+{\lambda}^{2}\,{y}_{c}^{2},\:\:\:\:\:\:\: \tau_{3}=\frac{\eta_{c}\,\lambda\,{y}_{c}^{2}}{2},
 \end{equation}with $y_{c}$ given in Eq. \eqref{37}. The eigenvalue $\tau_{1}$ is  always negative since $x_{c}^{2}\leq1$. In regard to $\tau_{2}$, it is always negative provided that $\gamma\geq1$. 
 On the other hand, the eigenvalue $\tau_{3}$ is negative if $\eta_{c}\,\lambda<0$.
 So, for $\hat{\xi}>0$ with $\gamma\geq1$, $\lambda>0$ and $\eta_{c}<0$ (or $\lambda<0$ and $\eta_{c}>0$), then point II.c is a stable node.
 
Therefore, point II.c is a late-time attractor and a viable cosmological solution (scalar-field dominant solution) with accelerated expansion. Unlike the late-time attractors I.d and I.e for constant $\alpha$, in this case the universe can enters in the scaling solutions I.b or I.c ( phase $\phi$MDE) with constant $\alpha$ and eventually approaches the late-time attractor II.c for dynamically changing $\alpha$, since in this case we can have $0<\hat{\xi}<1$ depending on the value of $\lambda$ in \eqref{39}.
In Fig. 1 we show the case when the system approaches the fixed point II.c with $\gamma=1$ (non-relativistic dark matter), $\lambda=0.6$, $\xi=-3\times10^{-3}$, and following Ref. \cite{13}, by way of example we consider the function $\alpha(u)=u_{c}-u$ with $u_{c}=1$ and $\eta_{c}=-1$. In this case $\Omega_{\phi}$ grows to 0.7 at the present epoch $N'\approx4$ and the system asymptotically evolves toward the values $\Omega_{\phi}=1$, $\Omega_{m}=0$ and $\omega_{\phi}=\omega_{eff}=-0.89<-1/3$.
Also, the universe undergoes a phase $\phi$MDE (scaling solution I.c) with $\Omega_{\phi}=1-\hat{\xi}\approx0.04$ and $\omega_{\phi}=\omega_{eff}=0$, before entering the late time attractor II.c. 

\begin{figure}[ht]
\centering
\includegraphics[width=0.65\textwidth]{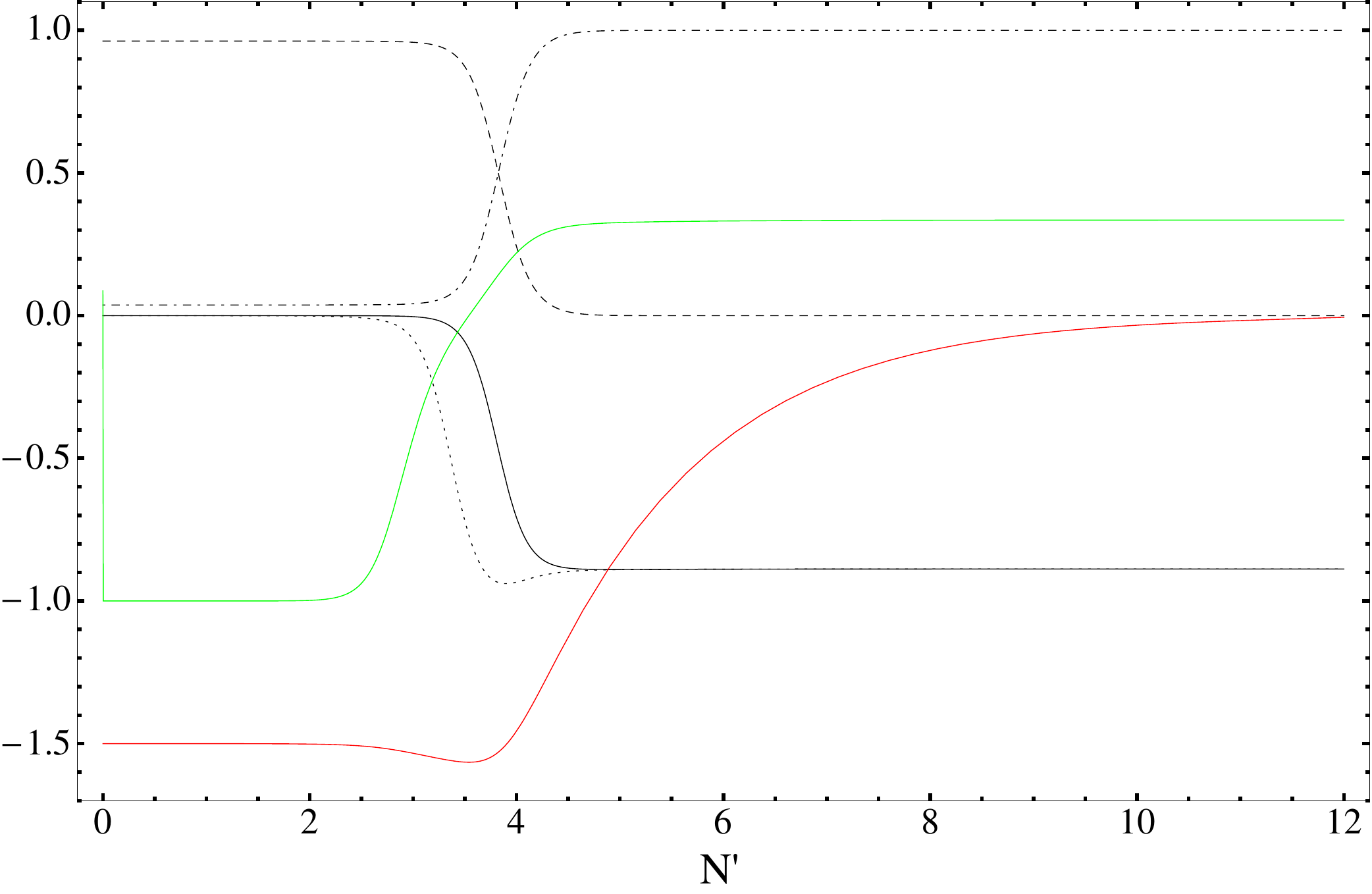}
\caption{Evolution of $\Omega_{m}$ (dashed), $\Omega_{\phi}$ (dotdashed), $\omega_{\phi}$ (dotted), $\omega_{eff}$ (solid), $x$ (green line, ending at $x_{c}\approx0.34$) and $\alpha(u)$ (red line, starting at $\alpha=-1.5$) with $\gamma=1$, $\lambda=0.6$ and $\xi\approx-3\times10^{-3}$. We choose initial
conditions $x_{i}=0.1$, $y_{i}=1.7\times10^{-6}$ and $u_{i}=2.5$ and by way of example we consider the function $\alpha(u)=u_{c}-u$ with $u_{c}=1$ and $\eta_{c}=-1$. The universe exits from scaling solution I.c with constant $\alpha=-1.5$, $\Omega_{\phi}\approx0.04$, $\omega_{\phi}=\omega_{eff}=0$ and approaches the late-time attractor II.c for dynamically changing $\alpha(u)$ with $\Omega_{\phi}\approx0.7$, $\Omega_{m}\approx0.3$ and accelerated expansion at the present epoch $N'\approx4$. The system asymptotically evolves toward the scalar-field dominant solution II.c with values $\Omega_{\phi}=1$, $\Omega_{m}=0$ and $\omega_{\phi}=\omega_{eff}=-0.89$.}
\label{figure}
\end{figure}

\section{Concluding remarks}

In Ref.~\cite{23} it was proposed a non-minimal coupling between a non-canonical scalar field (tachyon field) in the context of teleparallel gravity.
Here, by studying the dynamics of this tachyonic teleparallel dark energy model, we have found that, unlike teleparallel dark energy, in tachyonic teleparallel dark energy it is possible to have a phase $\phi$MDE, represented by the scaling solutions I.b and I.c of Table 1, which have some portions of the energy density of $\phi$ in the matter dominated era. 
The presence of this phase provides a distinguishable feature for matter density perturbations, as is the case of coupled dark energy in GR (see Refs. \cite{1, 4, 31}). However, in order to allow the universe to enter the phase $\phi$MDE, and then to fall within a viable cosmologically late-time attractor with accelerated expansion, it is necessary that the non-minimal coupling be ruled by a dynamically changing coefficient $\alpha(\phi)\equiv f_{,\phi}/\sqrt{f}$ , with $f(\phi)$ an arbitrary function of the scalar field $\phi$.
Following Ref.~\cite{13}, we considered then that $\alpha(\phi)$ can be expressed in terms of the dimensionless parameter $u\equiv\kappa \sqrt{f(\phi)}$, such that $\alpha(u)\rightarrow\alpha(u_{c})=0$, with $(x_{c}, y_{c}, u_{c})$ a fixed point of the system. We have found the fixed points (see Table 3) that are non-minimal generalization of the fixed points presented in Ref.~\cite{1} for tachyon field in GR. The scalar-field dominant solution II.c is a late-time attractor with accelerated expansion, and $\omega_{\phi}$ agrees with the observations. Also, it is possible in this case that the universe 
enters in the scaling solutions I.b or I.c (phase $\phi$MDE) for constant $\alpha$ and eventually approaches the late-time attractor II.c  with accelerated expansion for dynamically changing $\alpha(u)$, as can be seen in Fig 1. 

It should be noted that the formation of caustics in the field profile in the mass free space, for tachyon systems (Dirac-Born-Infeld systems)
is an undesirable feature as it indicates the failure of physical theories to explain the evolution of the field in that particular region \cite{A. Starobinsky, M. Sami}. As was shown in \cite{M. Sami}, in the FRW expanding Universe the caustic formation in tachyon systems takes place for 
potentials decaying faster than $1/\varphi^{2}$ at infinity (for the untransformed field $\varphi$), where the dust-like solution is a late time attractor of the dynamics. On the other hand, in the case of inverse power-law potentials, $V(\varphi)=V_{0}/\varphi^{n}$, $0\leq n\leq2$, 
dark energy is a late time attractor of dynamics and they are free of caustics \cite{M. Sami}. They may, therefore, be suitable for explaining the late time cosmic acceleration. So, since in the case of the model discussed, dark energy is a late time
attractor of the dynamics, which gives rise to cosmic repulsion that compete with the tendency of caustic formation, we expect the model to be free of caustics and multivalued regions in the field profile.

\section{Acknowledgments}

The author would like to thank J. G. Pereira for useful discussions and suggestions. He would like to thank also CAPES for financial support.

\end{document}